\documentclass[prd,twocolumn,a4paper,showpacs]{revtex4}

\usepackage{hyperref}
\usepackage{color,graphicx}
\usepackage{amsmath}

\begin{document}

\title{Forecasting neutrino masses from combining KATRIN and the CMB:\\ Frequentist and Bayesian analyses}
\date{September 7, 2007}
\author{Ole Host}
\affiliation{Dark Cosmology Centre, Niels Bohr Institute, University of Copenhagen, Juliane Maries Vej 30, DK-2100 Copenhagen, Denmark}
\author{Ofer Lahav, Filipe B.~Abdalla}
\affiliation{Department of Physics and Astronomy, University College London, Gower St., London WC1E 6BT, UK}
\author{Klaus Eitel}
\affiliation{Forschungszentrum Karlsruhe, Institut f\"ur Kernphysik, Postfach 3640, 76021 Karlsruhe, Germany} 

\begin{abstract}
We present a showcase for deriving bounds on the neutrino masses from laboratory experiments and cosmological observations. We compare the frequentist and Bayesian bounds on the effective electron neutrino mass $m_\beta$ which the KATRIN neutrino mass experiment is expected to obtain, using both an analytical likelihood function and Monte Carlo simulations of KATRIN. Assuming a uniform prior in $m_\beta$, we find that a null result yields an upper bound of about $0.17$ eV at 90\% confidence in the Bayesian analysis, to be compared with the frequentist KATRIN reference value of $0.20$ eV. This is a significant difference when judged relative to the systematic and statistical uncertainties of the experiment. On the other hand, an input $m_\beta=0.35$ eV, which is the KATRIN $5\sigma$ detection threshold, would be detected at virtually the same level. Finally, we combine the simulated KATRIN results with cosmological data in the form of present (post-WMAP) and future (simulated Planck) observations. If an input of $m_\beta=0.2$ eV is assumed in our simulations, KATRIN alone excludes a zero neutrino mass at $2.2\sigma$. Adding Planck data increases the probability of detection to a median $2.7\sigma$. The analysis highlights the importance of combining cosmological and laboratory data on an equal footing. 
\end{abstract}

\maketitle

One of the most important recent discoveries in physics is that neutrinos have non-zero masses. Since neutrinos are explicitly massless in the Standard Model, massive neutrinos necessarily involve new physics and may point the way for other developments. Therefore there is much need for improved knowledge of neutrino masses from both laboratory experiments and cosmological observations.

The fact that neutrinos are massive is demonstrated by atmospheric \cite{Ashie:2004mr,Sanchez:2003rb}, solar \cite{Aharmim:2005gt,Abdurashitov:2002nt,Bahcall:2000nu,Davis:1994jw,Altmann:2005ix}, reactor \cite{Apollonio:2002gd,Araki:2004mb,Boehm:2001ik} and accelerator \cite{Michael:2006rx,Ahn:2002up,Aliu:2004sq} neutrino oscillation experiments. Additionally, the masses are constrained from above by tritium beta decay \cite{Lobashev:2003kt,Kraus:2004zw}, neutrinoless double beta decay \cite{Aalseth:2000ud,Arnaboldi:2005cg,Arnold:2005rz,Klapdor-Kleingrothaus:2000sn}, and from bounds on the amount of hot dark matter in the Universe (see below for references). Since the oscillation experiments are only sensitive to square mass differences, the absolute neutrino mass scale has not been determined, apart from the disputed observation of neutrinoless double beta decay in the Heidelberg-Moscow experiment \cite{Klapdor-Kleingrothaus:2004wj,Aalseth:2002dt,Klapdor-Kleingrothaus:2006ff}. However, recent cosmological observations of the Cosmic Microwave Background \cite{Kuo:2002ua,Leitch:2004gd,MacTavish:2005yk,Sievers:2005gj,Spergel:2006hy}, the Large Scale Structure of the Universe \cite{Cole:2005sx,Tegmark:2003uf}, the Baryon Acoustic Oscillations in the matter power spectrum \cite{Eisenstein:2005su}, Type Ia Supernovae \cite{Astier:2005qq,Riess:2006fw,Wood-Vasey:2007jb}, and Lyman-$\alpha$ clouds \cite{McDonald:2004eu} show substantial evidence for the `standard' $\Lambda$CDM cosmological model. In that model case, the upper bound on the sum of neutrino masses $\sum m_\nu$ is driven well into the sub-eV region \cite{Elgaroy:2006ii,Fogli:2006yq,Fukugita:2006rm,Goobar:2006xz,Kristiansen:2006ky,Seljak:2006bg,Spergel:2006hy}. If so, it seems unlikely that a detection of the electron neutrino mass $m_\beta$ will be made at KATRIN \cite{Drexlin:2005zt}, the next generation tritium beta decay experiment which has a discovery potential of $0.35$ eV. However, there are two important issues with the cosmological data. Firstly, the results depend on which cosmological model is used, and different models contain different numbers of parameters. Adding more parameters will in general relax the bounds on any single parameter. Secondly, it should be noted that most of the predicting power relies on Lyman-$\alpha$ data and it is extremely hard to model these systems \cite{Seljak:2006bg}. This makes it challenging to assess the systematic effects associated with this measurement. Without Lyman-$\alpha$, current limits are in the region $0.6$--$0.7$ eV for $\sum m_\nu$. Hence a cleaner probe such as laboratory experiments, galaxy surveys or weak lensing surveys is very desirable. It should be noted that the next generation of large-scale surveys should be able to access the range down to $0.2$ eV for the sum of the neutrino masses \cite{Abdalla:2007ut}.

Comparison of various neutrino mass bounds is complicated by the fact that there are several different, but interdependent, observables involved (see, e.g., Fogli et al.~\cite{Fogli:2006yq}), but also by the use of different statistical methods which potentially leads to inconsistencies. The cosmology community usually adheres to Bayesian inference while experimental neutrino physicists tend to employ frequentist methods in oscillation experiments and absolute mass searches. 

In this paper we address these complications by comparing frequentist and Bayesian neutrino mass bounds in relation to the near-future KATRIN experiment and by combining the estimated KATRIN results with present and future cosmological observations in a consistent manner. First, we introduce the KATRIN experiment and then we compare the frequentist and Bayesian confidence intervals obtained from an approximate KATRIN likelihood function. Then we simulate KATRIN using Monte Carlo methods and compare a Bayesian analysis of our simulations with the frequentist analysis by the KATRIN collaboration. Finally, we predict the future neutrino mass sensitivity by combining our results with present (post-WMAP) and future (Planck) cosmological observations in a consistent Bayesian framework.

\section{The KATRIN experiment}
The Karlsruhe Tritium Neutrino experiment KATRIN is a next generation
single beta decay experiment investigating the kinematics of the tritium
beta spectrum near its kinematical endpoint \cite{Drexlin:2005zt}. It will measure the effective electron
neutrino mass through the nuclear beta decay of tritium
\begin{equation}
^{3}\text{H}\rightarrow\,^{3}\text{He}^++\mathbf{e}^-+\bar{\nu}_e.
\end{equation}
with an endpoint energy $E_0\simeq18.6$\,keV. Any non-zero (effective) mass of the electron antineutrino would lead to an end of the beta spectrum lower than the endpoint energy $E_0$ as well as to a distortion of the spectrum in the region near the endpoint.

KATRIN investigates the $\beta$-electron energy with an electrostatic
high-pass filter which measures the integrated spectrum of $\beta$-electrons near the
endpoint of the spectrum. The lower bound of the filter is determined
by a retarding potential $U$ and the number of electrons in a given
potential energy bin is \cite{kdr}
\begin{equation}\label{eq:nqu}
N(qU)=A \int_{0}^{E_0}\text{d}E\,\frac{\text{d}N}{\text{d}{E}}(E,E_0,m_\beta^2)\,
f_{\text{res}}(qU,E)+N_b,
\end{equation}
where $f_{\text{res}}$ is the experimental response function, $E_0$ is
the endpoint energy of the $\beta$-spectrum, $N_b$ is the background,
and $A$ is the signal amplitude of the spectrum. The differential
electron spectrum close to the endpoint is, approximately,
\begin{equation}\label{eq:dnde}
\frac{\text{d}N}{\text{d}{E}}\simeq C\,(E_0-E)\sqrt{(E_0-E)^2-m_\beta^2}.
\end{equation}
The observable is therefore the square of the effective mass. The aim of
the KATRIN experiment is a sensitivity of $m_\beta<0.2$ eV at 90\% confidence
in case of a null result or a $5\sigma$ discovery potential for $m_\beta\ge 0.35$ eV \cite{kdr}. It has been shown \cite{Bonn:2007su} that non-SM contributions to the beta decay such as right-handed currents have little impact on the neutrino mass measurement.

KATRIN is actually being built at the site of Forschungszentrum Karlsruhe,
having major components such as the main spectrometer vessel with its diameter of 10m and a length of 24m already installed. Commissioning of the 70m-long full setup comprising the windowless gaseous tritium source, the electron transport section, a tandem spectrometer and a semiconductor detector array is expected for 2010. To reach
the expected sensitivity, a measuring time of 3 full-beam-years is anticipated,
starting end of 2010.

Here, our interest in KATRIN is mainly its expected statistical performance.
The benchmark is the frequentist Monte Carlo analysis by the KATRIN
collaboration as given in the Design Report \cite{kdr}: The outcome is an
expected distribution of occurrence which is very close to Gaussian in
$m_\beta^2$ with a width of
$\sigma_\text{stat.}(m_\beta^2)=0.018\text{ eV}^2$. This is combined with the estimate of the total systematic uncertainty,
$\sigma_\text{sys.}(m_\beta^2)\le 0.017\text{ eV}^2$, which leads to the total uncertainty of $\sigma_\text{tot.}(m_\beta^2)=(\sigma_\text{stat.}^2+\sigma_\text{sys.}^2)^{1/2}=0.025$ eV$^2$. Taking the Gaussian shape of the frequency distribution, the 90\% confidence sensitivity on $m_\beta$ is then derived via 
\begin{equation}
L(90\%)=\sqrt{1.64\sigma_\text{tot.}},
\end{equation}
and the detection threshold is $m_\beta=[5\sigma_\text{tot.}(m_\beta^2)]^{1/2}=0.35$ eV.

\section{Bayesian vs.~frequentist confidence intervals}\label{se:cl}
Next we discuss the frequentist and Bayesian approaches to parameter inference and calculation of upper and lower bounds. 

Both frequentist confidence intervals and Bayesian credible intervals are constructed from a likelihood function $L(x,s)$ which is a function of the data $x$ and the parameter to be inferred $s$ (For simplicity, we consider only one parameter in this discussion). A frequentist confidence interval is a statement about frequency of occurrence. The interval $[s_1,s_2]$ is a member of the set of all intervals that fulfil the condition that the probability
\begin{equation}
P(s\in [s_1,s_2])=1-\alpha,
\end{equation}
for all $s$, in particular the unknown true value. Here, $1-\alpha$ is the stated confidence level of the interval. The values $s_1$ and $s_2$ are determined by the likelihood function. This defining property known as `coverage' ensures that, in the limit of a large ensemble of intervals, a fraction $1-\alpha$ of the ensemble will contain the true value.

In the unified approach of Feldman and Cousins \cite{Feldman:1997qc}, intervals are constructed from a likelihood ordering method which involves two steps. Firstly, for a fixed choice of $s$ a physically allowed value of $x$ is admitted to an `acceptance region' if the likelihood at that value is greater than at any other value not included in the interval already. In the continuous case, this means that the acceptance region defined by $[x_1,x_2]$ will satisfy
\begin{equation}
\int_{x_1}^{x_2}L(x,s)\mathrm{d}x=1-\alpha,
\end{equation}
so that, additionally, the likelihood function $L$ is everywhere greater inside the region than outside. Secondly, this procedure is repeated for every $s$ on a finely spaced grid. Having measured the best-fit value $x_0$, the confidence interval is the union of all the values of $s$ that include $x_0$ in their acceptance region.

On the other hand, the Bayesian credible interval is a statement about degree of belief. It attaches a probability to a range of values, indicating quantitatively how much it is believed that the actual value is within that range. Following Bayes' theorem, the posterior probability density in the true value of the parameter $s$ is 
\begin{equation}
P(s|x)=\frac{L(x|s)\pi(s)}{P(x)},
\end{equation}
where $L(x|s)$ is the likelihood function, which is now viewed as the conditional probability of the data given $s$, and $\pi(s)$ is the assumed prior probability density of the parameter. The normalizing factor $P(x)$, also known as the Bayesian evidence, is simply the integral of the numerator. The credible interval $[s_1,s_2]$ is calculated by integrating the posterior probability density so that
\begin{equation}\label{eq:ba}
\frac{1}{P(x)}\int_{s_1}^{s_2}L(x|s)\pi(s)\mathrm{d}s=1-\alpha.
\end{equation}
The framework can straightforwardly be generalized to multiple parameters, including nuisance parameters which are easy to handle. One can just integrate (marginalize) the multidimensional posterior distribution over the nuisance parameters to obtain the posterior in the parameter(s) of interest. The prior is a means of incorporating known information about $s$, which is then updated by the measurements in the form of the likelihood. One can think of at least three bases for the choice of prior: a known experimental result, a theoretical analysis, or `ignorance' about the value of the parameter.

\section{Comparison of intervals for KATRIN}
Now we study the frequentist and Bayesian intervals constructed from the predicted KATRIN likelihood function for a vanishing mass. As discussed, the observable is the square of the electron neutrino mass $m_\beta^2$ and for the purposes of fitting it is necessary to also consider negative values of this parameter, i.e.~$m_\beta^2<0$. The realness is then imposed afterwards. Therefore we assume a likelihood which is Gaussian in $m^2_\beta$ and has width $\sigma_\text{tot.}(m_\beta^2)=0.025$ eV$^2$, the estimated total error of KATRIN \cite{kdr}. We consider varying best-fit values $m^2_{\text{b.f.}}$ in the range $\pm5\sigma=\pm0.125$ eV$^2$ around zero. We calculate the frequentist confidence intervals in the unified approach of Feldman \& Cousins \cite{Feldman:1997qc} but we state the bounds in $m_\beta$, the parameter of interest, rather than in $m_\beta^2$. We also calculate Bayesian credible intervals for three different priors which are constant in $m_\beta^2$, $m_\beta$, and $\log m_\beta$, respectively. The first two priors are bounded from below at zero mass and the logarithmic prior is bounded at $7$ meV, suggestive of lower limits from oscillation data. All the priors are bounded from above at $1$ eV. 

There is a certain freedom in calculating the Bayesian intervals, since an interval containing a specific probability mass is not uniquely determined without further constraints. We choose to use the shortest possible interval, integrated in the metric in which the chosen prior is constant. However, we stress that this choice is not unique.
\begin{figure}
\includegraphics[width=\columnwidth]{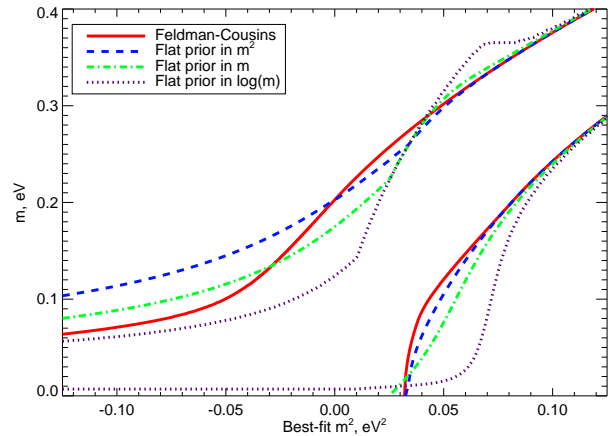}
\caption{Upper and lower bounds at 90\% confidence as a function of the best-fit value calculated in the frequentist Feldman-Cousins approach, and in the Bayesian approach with three different priors. The interpretation of a null result, $m_\text{b.f.}^2=0$ is very sensitive to the choice of analysis, while the confidence interval for a frequentist $5\sigma$ detection at $m_\text{b.f.}^2=0.125\text{ eV}^2$ is virtually the same in all cases. A Gaussian likelihood function in $m_\beta^2$ is assumed. Note that, for the purpose of fitting, the observable $m_\beta^2$ can take on both positive and negative values.}
\label{fi:cls}
\end{figure}

The resulting upper and lower bounds at 90\% confidence are shown in Fig.~\ref{fi:cls}. The intervals, regardless of method and assumed prior, approximately converge when the best-fit value is physical and far from the boundary, i.e.~for $m_\text{b.f.}^2\gtrsim0.125\text{ eV}^2=5\sigma$. The behaviour for $m_\text{b.f.}^2\lesssim0$ varies considerably; in particular the Feldman-Cousins upper bound drops faster than any Bayesian upper bound when going from positive to negative $m_\text{b.f.}^2$. Also, the Bayesian upper bounds, except for the logarithmic prior, are more conservative for very negative best-fits. The behaviour of the logarithmic prior is distinct due to the larger prior probability density attached to small values of $m_\beta$. This causes the logarithmic prior to yield the tightest upper bounds for small $m^2_\text{b.f.}$. At first glance the logarithmic prior, or Jeffreys' prior \cite{19321001} as it is also known, may appear artificial but it has the nice property that it treats all powers of the parameter equally. Also, as shown by Jaynes \cite{jaynes}, it is the prior which contains the least information about a scale parameter, i.e.~an `ignorance' prior.

Finally, we note that the question of whether a nonzero neutrino mass is detected can also be discussed from a model selection perspective. In the frequentist framework this is the rejection of the null hypothesis that the mass is zero, while in the Bayesian sense one may compare the Bayesian evidence for models with and without the neutrino mass \cite{Trotta:2005ar}. 

\section{Bayesian bounds from simulations of KATRIN}\label{se:bb}
In order to understand better the statistical perfomance of KATRIN from a Bayesian point of view we perform an independent Monte Carlo simulation of the experiment which is analyzed in the Bayesian framework and compared with reference results.

The model is essentially given by eqs.~\eqref{eq:nqu} and \eqref{eq:dnde}. There are four input parameters: $m_\beta^2$, the endpoint energy of the beta spectrum $E_0$, the background per bin $N_b$, which is assumed to be independent of $U$, and the signal amplitude of the spectrum, $A$. We adjust $A$ to reproduce the statistical error of the frequentist KATRIN simulations. We also recover the approximately Gaussian shape of the likelihood function.

We assume priors which are uniform in each of the nuisance parameters and in $m_\beta$. For each realization of the data we calculate the likelihood function on a grid in the four dimensional parameter space which extends more than five standard deviations to each side of the input values. The grid contains 50 points along each of the three nuisance parameters and 130 points along the $m_\beta^2$ direction in the range $[-0.1,1] $ eV$^2$ and it is constructed so that it is regular in $m_\beta$. From this tabulated likelihood function we marginalize over the three nuisance parameters and convolve the result with a Gaussian with width $\sigma_{\mathrm{sys}}(m_\beta^2)=0.017$ eV$^2$ to model the maximum expected systematic error \cite{kdr}. The result is the posterior probability density of $m_\beta$ from which the Bayesian credible interval can be calculated as in section \ref{se:cl}.

We have checked that the resolution of the grid is sufficient by calculating the total probability contained within a given interval for ten datasets resolved on our grid and on a reference grid with $100$ points for each nuisance parameter. The results differ by $\mathcal{O}(10^{-5})$.

\begin{figure}[tbp]
\includegraphics[width=\columnwidth]{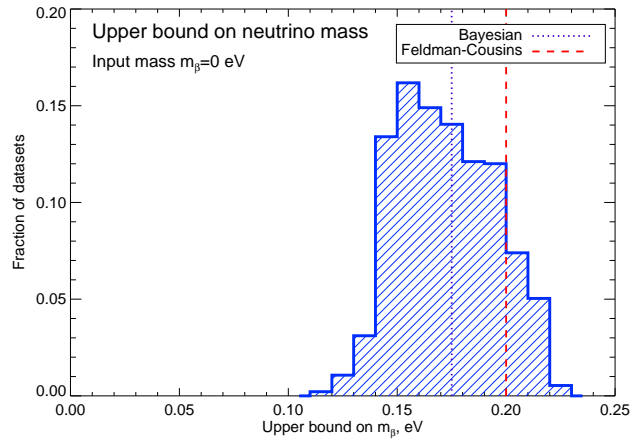}
\caption{Distribution of 90\% upper bounds on $m_\beta$ for an input $m_\beta=0$, obtained in the Bayesian analysis assuming a uniform prior in $m_\beta$. As expected, the distribution is scattered around the upper limit from Fig.~\ref{fi:cls} for a Gaussian likelihood with a uniform prior in $m_\beta$, which is significantly lower than the frequentist expectation. Out of 1000 simulated datasets, 67 did not include zero at 90\% confidence.}
\label{fi:cz}
\end{figure}

\begin{figure}[tbp]
\includegraphics[width=\columnwidth]{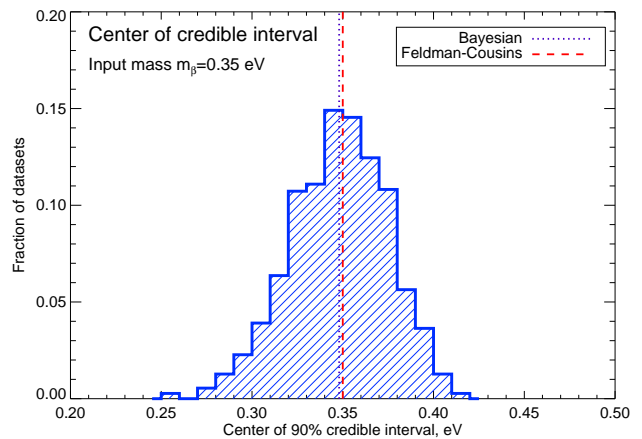}
\caption{Distribution of central values of 90\% confidence intervals for an input $m_\beta=0.35$ eV. As in Fig.~\ref{fi:cz}, the distribution is scattered around the central value for the corresponding interval for a uniform prior in Fig.~\ref{fi:cls} but in this case there is no discrepancy with the frequentist expectation for the central value.}
\label{fi:cnz}
\end{figure}

We consider two cases with input masses $m_\beta=0$ and $m_\beta=0.35$ eV, respectively. For the zero input we calculate the upper bound and for the nonzero input the central value of the confidence interval, both at 90\% confidence. The resulting distributions are shown in Figs.~\ref{fi:cz} and \ref{fi:cnz}, respectively. Also shown are the expected upper bounds/central values (dotted lines) obtained in section \ref{se:cl} as well as the frequentist expectation.

The mean upper bound of the distribution in Fig.~\ref{fi:cz} is $0.17$ eV so it is clear that the Bayesian analysis is more optimistic than the frequentist which yields the bound $0.20$ eV. This is in agreement with the analysis in section \ref{se:cl}, in particular with the upper bound of $0.175$ eV obtained for a uniform prior in $m_\beta$ and best-fit $m^2_\text{b.f.}=0$, as shown in Fig.~\ref{fi:cls}. This difference between the frequentist and Bayesian upper bound is large since, if the frequentist analysis were to yield an upper bound of 0.17 eV, the total error of $\sigma_\text{tot.}(m_\beta^2)$ would need to be reduced by 40\%.

\begin{figure}[tbp]
\includegraphics[width=\columnwidth]{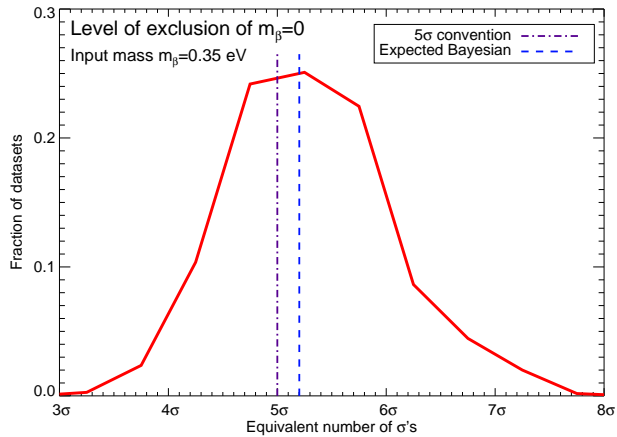}
\caption{The level at which $m_\beta=0$ is just excluded, given in terms of the equivalent number of $\sigma$'s for the input $m_\beta=0.35$ eV. The left vertical line is the conventional $5\sigma$ detection threshold and the 63\% of the datasets to the right of this line exclude $m_\beta=0$ at more than $5\sigma$. The right vertical line is the expected exclusion level in the Bayesian analysis, assuming a uniform prior in $m_\beta$ and a Gaussian likelihood in $m_\beta^2$. 51\% of the datasets are to the right of this line.}
\label{fi:ex}
\end{figure}

For the input $m_\beta=0.35$ eV in Fig.~\ref{fi:cnz}, the central value distribution is centered on the input and the Bayesian and frequentist analyses yield the same result. Again, this is in agreement with the intervals in Fig.~\ref{fi:cls} since the input mass is much greater than zero. The crucial question for the discovery potential, however, is at which confidence level the simulated datasets just exclude $m_\beta=0$, in terms of the equivalent number of $\sigma$'s. This is shown in Fig.~\ref{fi:ex} in terms of the equivalent number of $\sigma$'s. Also indicated is the conventional $5\sigma$ discovery limit. It can be seen that a majority of the datasets exclude $m_\beta=0$ at a greater confidence level than $5\sigma$. The reason for this is mainly the choice of prior; a uniform prior in $m_\beta^2$ instead would only show random scatter around the $5\sigma$ line. When assuming a likelihood function which is Gaussian in $m_\beta^2$ and adopting a uniform prior in $m_\beta$, $m_\beta=0$ is excluded at a probability corresponding to $5.2\sigma$. This value is very close to the median of the distribution in Fig.~\ref{fi:ex}.

From a frequentist point of view Fig.~\ref{fi:ex} indicates that the Bayesian credible intervals with uniform prior in $m_\beta$ are more likely to undercover. This means that less than the fraction $1-\alpha$ of an ensemble of intervals will contain the true value of the parameter---at least at this extreme confidence level.

In summary, the Bayesian analysis carried out here is more optimistic than the frequentist in this respect in the sense that both the sensitivity and discovery potential are lower.

\section{Combination with cosmology}
Having analyzed the differences between mass bounds derived by the different statistical methods, we proceed by combining our KATRIN simulation with present and future cosmological observations in the Bayesian framework. 

\begin{figure}
\includegraphics[width=\columnwidth]{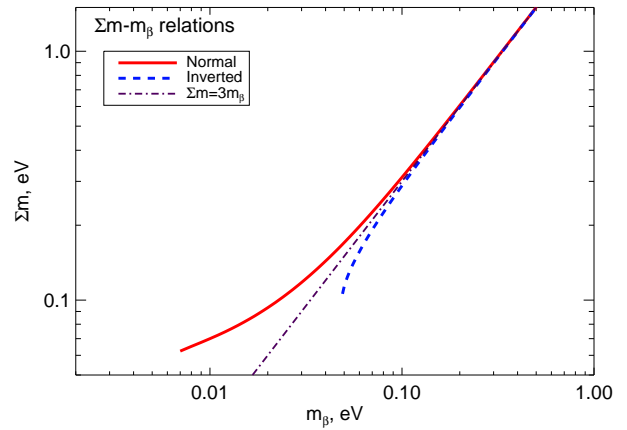}
\caption{Relation between $m_\beta$ and $\sum m_\nu$ for normal and inverted hierarchies, and the naive relation $\sum m_\nu=3m_\beta$. For $m_\beta \gtrsim 0.1$ eV there is virtually no differences between the relations.}
\label{fi:osc}
\end{figure}

The observable in tritium beta decay $m_\beta$ and the cosmological observable $\sum m_\nu$, are related through the neutrino mixing angles and square mass differences which are probed by oscillation experiments. The mixing angles form a parametrization of the PMNS matrix which relates the weak interaction neutrino eigenstates $\left|\nu_\alpha\right\rangle$ to the mass eigenstates $\left|\nu_i\right\rangle$ through $\left| \nu_\alpha \right\rangle = \sum_i U_{\alpha i}^\ast \left| \nu_i \right\rangle$ \cite{PDBook}. We simplify the analysis by fixing the oscillation data \cite{Araki:2004mb,Michael:2006rx,Apollonio:2002gd} at best-fit values. In particular, we assume $\sin\theta_{13}=0$ which determines $m_\beta=(m_1 +m_2)/2$, and the masses can then be written in terms of observables as
\begin{subequations}\label{eq:msq}\begin{eqnarray}
m_1^2 &=& m_\beta^2 -\frac{1}{2}\Delta m_{12}^2, \\
m_2^2 &=& m_\beta^2 +\frac{1}{2}\Delta m_{12}^2,\\
m_3^2 &=& m_\beta^2 +\frac{1}{2}\Delta m_{12}^2\pm\Delta m_{23}^2.\label{eq:m3}
\end{eqnarray}
\end{subequations}
The only freedom left is the hierarchy, i.e.~the sign of $\Delta m_{23}^2$ in \eqref{eq:m3}. The best-fit square mass differences are
\begin{subequations}\begin{eqnarray}
\Delta m_{12}^2&=&7.9\times10^{-5}\text{ eV}^2,\\
\Delta m_{23}^2&=&2.6\times10^{-3}\text{ eV}^2.
\end{eqnarray}\end{subequations}
The hierarchy determines if the mass $m_3$ is greater or lesser than the pair $m_1$ and $m_2$. Relations between $m_\beta$ and $\sum m_\nu$ are shown in Fig.~\ref{fi:osc}.

In principle, the uncertainties on the neutrino mixing angles and square mass differences should be taken into account, but for the purpose of this showcase we deem the best-fit values sufficient. It was found in ref.~\cite{Fogli:2006yq} that the uncertainty in the $m_\beta$--$\sum m_\nu$ relation is very small, except for the smallest allowable masses.

We combine simulated KATRIN posteriors with present cosmological data in the form of the WMAP 3-year data to which data from large-scale structure surveys and type Ia supernovae are added. Additionally, we also combine the KATRIN results with a simulation of the Planck satellite, the next cosmic microwave background satellite.

\subsection{Bounds from the CMB}
\begin{figure}
\includegraphics[width=\columnwidth]{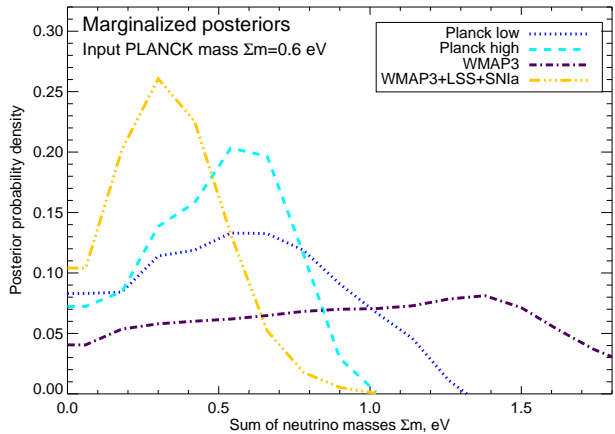}
\caption{The one-dimensional marginalized posteriors for $\sum m_\nu$ as obtained in in the simulation of Planck, computed with an input mass of input $\sum m_\nu=0.6$ eV, and the WMAP analyses. These posteriors yield upper bounds of $\sum m_\nu<1.1$ eV (Planck low), $\sum m_\nu<0.81$ eV (Planck high), $\sum m_\nu<1.8$ eV (WMAP only), and $\sum m_\nu < 0.66$ eV (WMAP+LSS+SNIa), all at 95\% confidence.}
\label{fi:cos}
\end{figure}

We include present limits on $\sum m_\nu$ from cosmology in the form of the analysis of the WMAP 3-year data, as performed by the WMAP team \cite{Spergel:2006hy}. We use their MCMC chains, available at \cite{lambda}, for a $\Lambda$CDM cosmology described by the six minimal parameters plus the neutrino energy density. We use the chains for the largest combined dataset, which includes WMAP and other CMB data as well as large scale structure and supernovae measurements. It was shown in ref.~\cite{Kristiansen:2006ky} that an identified problem with the WMAP likelihood code does not affect neutrino mass bounds from combined datasets. On its own, this data set yields an upper bound on the sum of the neutrino masses of $\sum m_\nu < 0.66$ eV at 95\% confidence. This strong bound is mainly caused by the large-scale structure and small-scale CMB data included; the WMAP data alone yield the much weaker constraint $\sum m_\nu < 1.8$ eV, also at 95\% confidence \cite{Spergel:2006hy,Fukugita:2006rm}. 

Additionally, we have obtained forecasts from the Planck satellite by using the technique described in Abdalla \& Rawlings \cite{Abdalla:2007ut}. This consists mainly in assuming a fiducial model and calculating the expected rejection probability of a model around this fiducial value. We have assumed conservative values for the sensitivity by taking only one usable science channel with 8 detectors, a noise effective temperature of $120\text{ }\mu\text{K} \sqrt{\text{s}}$, an angular resolution of 10 arcseconds, and 60\% sky coverage in a one year survey. We have also considered a more optimistic forecast which would cover around 70\% of the sky in a two year survey. In both cases the input fiducial mass was taken as $\sum m_\nu=0.6$ eV and the other
cosmological parameters considered were: $\Omega_m=0.3$, $\Omega_b=0.04$,
$n_s=0.95$, $\Omega_\Lambda=0.7$, $n_{\text{run}}=0.00$, $\sigma_8=0.8$, $\tau =0.09$, and we assumed a flat Universe. The resulting posteriors yield upper bounds of $\sum m_\nu<1.1$ eV (conservative) and $\sum m_\nu<0.81$ eV (optimistic) at 95\% confidence.

Both the WMAP+LSS+SNIa posterior and the simulated Planck posteriors, marginalized over all parameters except $\sum m_\nu$, are shown in figure \ref{fi:cos}. The Planck simulation yields higher upper bounds on $\sum m_\nu$ since it is based on the CMB alone, and because the input to the simulation is already close to the upper bound derived from WMAP+LSS+SN.

\subsection{Combined KATRIN and CMB}
Now we combine our simulation of KATRIN with the cosmological observations discussed above.
\begin{figure}
\includegraphics[width=\columnwidth]{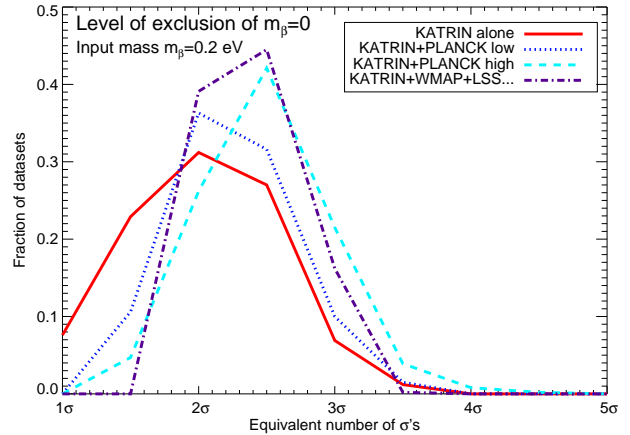}
\caption{Exclusion levels of $m_\beta=0$ given in the equivalent number of $\sigma$'s for an input $m_\beta=0.2$ eV to the KATRIN simulation, with and without cosmological data. The distribution for KATRIN alone differs from Fig.~\ref{fi:ex} only through the input. The extra data from cosmology clearly improve the detection threshold, although for this input mass a $4\sigma$ detection is still unlikely.}
\label{fi:excl}
\end{figure}

We simulate a new set of 1000 KATRIN experiments with an input mass of $m_\beta=0.2$ eV, the expected KATRIN 90\% upper bound on a null result. Using the relations in \eqref{eq:msq}, we convert each simulated $m_\beta$-posterior into a posterior in $\sum m_\nu$, assuming normal hierarchy for definiteness. For simplicity, we do not propagate the small errors from the oscillation data. The simulated KATRIN posteriors are combined with each of the cosmology posteriors individually and for each of these we calculate the confidence level at which a `vanishing' neutrino mass is excluded. There is a technical point here since converting the KATRIN observable using \eqref{eq:msq} implies that neutrinos are massive. In other words, whereas it is possible to examine a KATRIN posterior and a cosmology posterior in isolation and determine whether there is evidence for a nonzero neutrino mass in each posterior, the two can only be combined under the assumption that both $m_\beta$ and $\sum m_\nu$ are nonzero. Therefore we search for the confidence level at which the lower bound of the interval just reaches the lower limit of $\sum m_\nu$ and regard this as the detection threshold (for that posterior) of a massive neutrino. Otherwise the detection of the nonzero mass is from the oscillation data rather than the KATRIN+cosmology posterior.

The procedure yields distributions of the exclusion level of a `vanishing' mass for KATRIN alone and for the two types of combined posteriors, which are shown in Fig.~\ref{fi:excl}. For the KATRIN posteriors alone, the median detection level is $2.2\sigma$ and it is clear that a large fraction of the simulated experiments do not exclude $m_\beta=0$ even at $2\sigma$. On the other hand, most experiments do exclude $m_\beta=0$ at 90\% confidence, in agreement with expectations based on the analysis in section \ref{se:cl}. 

The addition of the cosmological data improves the detectability of a nonzero mass. The combination with WMAP+LSS+SNIa yields a median detection level of $2.6\sigma$ while the combination the Planck yields $2.4\sigma$ (conservative) and $2.7\sigma$ (optimistic). More importantly, the widths of the distributions become smaller so that many more combined posteriors exclude a vanishing mass at more than $2\sigma$. On the other hand, there are no detections at more than $4\sigma$ which is a consequence of the fact that neither of the three datasets are able to provide significant detections on their own. The combination of the high specs-Planck simulation with KATRIN actually performs slightly better than WMAP+LSS+SNIa and KATRIN, which is likely due to the fact that the WMAP+LSS+SNIa posterior is peaked at a much lower value of $\sum m_\nu$ than the input $0.6$ eV used in the simulations. In summary, the combination of KATRIN and cosmological data may well extend the discovery potential of the experiments into regions not accessible by either type of experiment alone. We note that additional large-scale structure data could improve the detection potential further.
 
\section{Conclusions}
We have presented a showcase for estimating the neutrino masses by combining data from laboratory and cosmological experiments. Specifically, we combine simulated results from KATRIN with cosmological probes. A key issue in this context is the consistent application of statistical methods to all data. We analyze the differing results obtained by the frequentist and Bayesian approaches, taking simulated KATRIN likelihoods as an example. Then we combine the KATRIN results with both the present cosmological constraints after WMAP and simulated future data from the Planck satellite. 

The comparison of frequentist and Bayesian methods shows important differences for KATRIN, where the likelihood is likely peaked close to the physical boundary $m_\beta=0$. While the KATRIN reference upper bound to a null result is $0.2$ eV at 90\% confidence, the Bayesian result is 0.17 eV, assuming a uniform prior, and even lower if a logarithmic prior is assumed. As explained in section \ref{se:bb}, this is a sizable difference judged against the systematic and statistical uncertainties. Far away from the zero mass boundary, the different approaches converge.

The combined constraints from laboratory and cosmological data are investigated by assuming an electron neutrino mass of 0.2 eV. In this case KATRIN alone will provide a $2.2\sigma$ detection, based on $10^3$ simulations of KATRIN. The $10^3$ exclusion levels span a rather wide range, from zero detection to about $4\sigma$. The addition of CMB data increases the detection level somewhat to $2.6\sigma$ (WMAP+LSS+SNIa) or $2.4\sigma-2.7\sigma$ (Planck) but, importantly, narrows the range of exclusion levels so that in most cases a $>2\sigma$ detection is found.

Our analysis highlights the impact of the choice of statistical analysis, and it demonstrates the application of Bayesian methods to combined datasets.
 
\begin{acknowledgments}
We thank Antony Lewis, Ruben Saakyan, and Roberto Trotta for useful comments. The Dark Cosmology Centre is funded by the Danish National Research Foundation.
\end{acknowledgments}

\bibliography{neutrino,stats}

\end{document}